\documentclass[a4paper,11pt]{article}
\usepackage{jheppub} % for details on the use of the package, please
\usepackage[T1]{fontenc} % if needed

\usepackage[latin1]{inputenc}
\usepackage[english]{babel}

\usepackage{amssymb,amsthm,cancel,hyperref,graphicx,xcolor}
\usepackage{picinpar,graphicx,xypic}
\usepackage{booktabs}
\usepackage{mathrsfs}
\usepackage{amsfonts}
\usepackage{latexsym}
\usepackage{booktabs,graphicx,hyperref,epsfig}
\usepackage{bm}
\allowdisplaybreaks
\def\bea{\begin{align}}
\def\eea{\end{align}}
\def\beq{\begin{equation}}
\def\eeq{\end{equation}}
\def\ba{\begin{eqnarray}}
\def\ea{\end{eqnarray}}
\def\be{\begin{equation}}
\def\ee{\end{equation}}

\def\({\left(}
\def\){\right)}
\def\[{\left[}
\def\]{\right]}

\def    \hepph  #1 {{\tt hep-ph/#1}}
\def    \hepex  #1 {{\tt hep-ex/#1}}
\long\def\symbolfootnote[#1]#2{\begingroup%
\def\thefootnote{\fnsymbol{footnote}}\footnote[#1]{#2}\endgroup}

\numberwithin{equation}{section}
\title{Addendum to: Constraints on the quartic Higgs self-coupling from double-Higgs production at future hadron colliders}

\author[a]{Wojciech Bizo\'{n},}
\author[b]{Ulrich Haisch,}
\author[c]{Luca Rottoli,}
\author[d,e]{Zach Gillis,}
\author[f]{Brian Moser}
\author[d,e,g]{and Philipp Windischhofer}

\affiliation[a]{Krakow, Poland}
\affiliation[b]{Max Planck Institute for Physics, F{\"o}hringer Ring 6,  80805 M{\"u}nchen, Germany}
\affiliation[c]{University of Zurich, Winterthurerstrasse  190, 8057 Zurich, Switzerland}
\affiliation[d]{Department of Physics, University of Chicago, Chicago, IL 60637, USA}
\affiliation[e]{Enrico Fermi Institute, University of Chicago, Chicago, IL 60637, USA}
\affiliation[f]{EP Department, CERN, 1211 Geneva 23, Switzerland}
\affiliation[g]{Kavli Institute for Cosmological Physics, University of Chicago, Chicago, IL 60637, USA}

\emailAdd{haisch@mpp.mpg.de}
\emailAdd{zachgillis@uchicago.edu}
\emailAdd{luca.rottoli@physik.uzh.ch}
\emailAdd{brian.moser@cern.ch}
\emailAdd{philipp.windischhofer@cern.ch}

\preprint{
\begin{flushright}
ZU-TH 10/24
\end{flushright}
}

\abstract{We study inclusive double-Higgs boson production at the LHC and at the HL-LHC including variations of the trilinear and of the quartic Higgs boson self-couplings at next-to-leading order (NLO) in QCD with full top quark mass dependence. Our results include the two-loop contributions to the $gg \rightarrow HH$ amplitudes that involve a modified $h_4$ vertex calculated in Ref.~\cite{Bizon:2018syu}. We present results at 13, 13.6 and 14 TeV centre-of-mass energies. The implementation of the calculation is made publicly available in the POWHEG-BOX-V2 Monte Carlo framework.}

\begin{document} 
\maketitle
\flushbottom

\section{Introduction}
\label{sec:intro}

The Higgs potential in the Standard Model (SM) of particle physics is at present still largely unexplored.
At hadron colliders, the shape of the Higgs potential can be determined only by measuring the Higgs boson self-couplings, which is particularly challenging due to the smallness of the cross-sections for double- and triple-Higgs boson production.
Recent searches at ATLAS~\cite{ATLAS:2022jtk} and CMS~\cite{CMS:2022dwd} for double-Higgs boson production with the full LHC Run 2 luminosity of $\sim 140$~fb$^{-1}$ provide upper limits on the measured $\sigma_{HH}$ cross section of order 3 times the SM value, mainly by combining the $b \bar b \gamma \gamma$, $b \bar b \tau^{+} \tau^{-}$ and $b \bar b b \bar b$ decay channels. 
These results translate into constraints on the observed coupling modifiers of the trilinear Higgs boson self-coupling modifier of $-0.4 < \kappa_3 < 6.3$, $-1.2 < \kappa_3 < 6.5 $ at ATLAS and CMS, respectively.
Although these constraints will improve during the High Luminosity phase of the LHC (HL-LHC) it is not yet certain whether double-Higgs boson production in the SM could be unequivocally observed.
Current estimates indicate that a $4 \sigma$ significance could be achieved for $\mathcal L \sim 3000$~fb$^{-1}$ when combining both experiments and all decay channels~\cite{Cepeda:2019klc}.
The possibility to obtain direct bounds on the quartic Higgs boson self-coupling modifier $\kappa_4$ from triple-Higgs boson production are even more uncertain due to its very small cross-section.
Although loose bounds on $\kappa_4$ may be obtained at the HL-LHC~\cite{Stylianou:2023xit}, even a 100 TeV hadron collider could only determine the SM rate to an accuracy of order one~\cite{Contino:2016spe,Mangano:2016jyj,Papaefstathiou:2015paa,Chen:2015gva,Kilian:2017nio,Fuks:2017zkg}.

In this context, it becomes important to explore complementary approaches to constrain the Higgs potential.
An alternative strategy is to determine the Higgs boson self-couplings indirectly.
This approach was initially proposed to constrain $\kappa_3$ via precise measurement of differential distribution of single Higgs boson production~\cite{McCullough:2013rea,Gorbahn:2016uoy,Degrassi:2016wml} and was later used to constrain the quartic Higgs boson self-coupling in Refs.~\cite{Bizon:2018syu,Borowka:2018pxx}.

%In this Addendum we revisit the calculation of Ref.~\cite{Bizon:2018syu} and provide results for inclusive double-Higgs production at the LHC and at the HL-LHC including the relevant electroweak (EW) two-loop amplitudes that involve a modified $h_4$ vertex.
In this Addendum we use the calculation of Ref.~\cite{Bizon:2018syu}, originally used to compute predictions for inclusive double-Higgs production at future hadron-hadron colliders, at the centre-of-mass energies relevant for the LHC and the HL-LHC.
Our results include the relevant electroweak (EW) two-loop amplitudes that involve a modified $h_4$ vertex whose calculation was presented in Ref.~\cite{Bizon:2018syu}.
Our predictions are supplemented by the next-to-leading order (NLO) QCD corrections with variations of the trilinear Higgs boson self-coupling with the full top quark mass dependence~\cite{Borowka:2016ehy,Borowka:2016ypz,Heinrich:2017kxx,Heinrich:2019bkc}.
The implementation of the above calculation, which allows for arbitrary variations of the trilinear and of the quartic Higgs boson self-couplings, is made publicly available in the POWHEG-BOX~\cite{Alioli:2010xd} Monte Carlo framework.

%
%
% can be written compactly as
%\begin{equation}
%	V = - \mu |H|^2 + \lambda |H|^4 
%\end{equation}
%where $H$ 

\section{Inclusive double-Higgs production at the LHC and the HL-LHC}
\label{sec:analytical}

In this section we report results for the inclusive production cross sections at the LHC and at the HL-LHC, considering centre-of-mass energies of 13 TeV, 13.6 TeV and 14 TeV.
The formul\ae\ have been obtained with the aforementioned POWHEG-BOX implementation of double-Higgs production at NLO QCD, using PDF4LHC15 NLO parton distribution functions~\cite{Butterworth:2015oua} through the LHAPDF interface~\cite{Buckley:2014ana}.
The implementation of the NLO QCD corrections is based on the latest version of the \texttt{ggHH} code that fixed an error in the two-loop amplitude which affected the results for $\kappa_3 \neq 1$ (see Refs.~\cite{Heinrich:2022idm,Bagnaschi:2023rbx}).
Our NLO predictions are obtained in the full theory using $m_{\mathrm{top}} = 173$~GeV that corresponds to the value hard-coded in the virtual matrix element computed in Ref.~\cite{Borowka:2016ehy}.
We display results for our central predictions and for the upper and lower values of the scale uncertainty envelope.
The central renormalisation and the factorisation scales are set to $\mu_R=\mu_F=m_{HH}/2$. 
The scale uncertainty is calculated via a canonical 7-scale variation envelope by varying $\mu_R$ and $\mu_F$ by a factor 2 with $ 1/2 \leq \mu_R/\mu_F \leq 2 $.

With $\Delta\kappa_3 = \kappa_3 - 1$ and $\Delta\kappa_4 = \kappa_4 - 1$, the inclusive production cross-sections take the form,
\begin{align}
    \sigma(pp \rightarrow hh)^{\rm central}_{13\, \rm TeV} &= 27.8\,\mathrm{fb}\times
             [1 - 0.874(\Delta\kappa_3) + 1.46\cdot 10^{-3}(\Delta\kappa_4)  + 0.333(\Delta\kappa_3)^2 \nonumber \\
        & + 7.91\cdot 10^{-4}(\Delta\kappa_3\Delta\kappa_4) + 2.71\cdot 10^{-5}(\Delta\kappa_4)^2  -1.60\cdot 10^{-3}(\Delta\kappa_3)^2(\Delta\kappa_4)   \nonumber \\
        &-1.89\cdot 10^{-5}(\Delta\kappa_3)(\Delta\kappa_4)^2 + 9.82\cdot 10^{-6}(\Delta\kappa_3)^2(\Delta\kappa_4)^2 ]\, ,
 \end{align}

\begin{align}
    \sigma(pp \rightarrow hh)^{\rm up}_{13\, \rm TeV} &= 31.6\,\mathrm{fb}\times
         [1 - 0.889(\Delta\kappa_3) + 1.42\cdot 10^{-3}(\Delta\kappa_4) 
         + 0.345(\Delta\kappa_3)^2 \nonumber \\
         & + 7.83\cdot 10^{-4}(\Delta\kappa_3\Delta\kappa_4) + 2.66\cdot 10^{-5}(\Delta\kappa_4)^2 -1.59\cdot 10^{-3}(\Delta\kappa_3)^2(\Delta\kappa_4)\nonumber \\ & -1.86\cdot 10^{-5}(\Delta\kappa_3)(\Delta\kappa_4)^2 
        + 9.72\cdot 10^{-6}(\Delta\kappa_3)^2(\Delta\kappa_4)^2 ]\, ,
    \end{align}

\begin{align}
    \sigma(pp \rightarrow hh)^{\rm low}_{13\, \rm TeV} &= 24.2\,\mathrm{fb}\times
         [1 - 0.864(\Delta\kappa_3) + 1.48\cdot 10^{-3}(\Delta\kappa_4)
        + 0.326(\Delta\kappa_3)^2 \nonumber \\ & + 8.02\cdot 10^{-4}(\Delta\kappa_3\Delta\kappa_4) + 2.75\cdot 10^{-5}(\Delta\kappa_4)^2  -1.62\cdot 10^{-3}(\Delta\kappa_3)^2(\Delta\kappa_4) \nonumber \\ & -1.92\cdot 10^{-5}(\Delta\kappa_3)(\Delta\kappa_4)^2
        + 9.95\cdot 10^{-6}(\Delta\kappa_3)^2(\Delta\kappa_4)^2 ]\, ,
    \end{align}

\begin{align}
    \sigma(pp \rightarrow hh)^{\rm central}_{13.6\, \rm TeV} &= 30.8\,\mathrm{fb}\times
         [1 - 0.870(\Delta\kappa_3) + 1.47\cdot 10^{-3}(\Delta\kappa_4)
        + 0.330(\Delta\kappa_3)^2 \nonumber \\ &
        + 7.84\cdot 10^{-4}(\Delta\kappa_3\Delta\kappa_4) + 2.72\cdot 10^{-5}(\Delta\kappa_4)^2 -1.58\cdot 10^{-3}(\Delta\kappa_3)^2(\Delta\kappa_4) \nonumber \\ & -1.90\cdot 10^{-5}(\Delta\kappa_3)(\Delta\kappa_4)^2
         + 9.77\cdot 10^{-6}(\Delta\kappa_3)^2(\Delta\kappa_4)^2 ]\, ,
    \end{align}

\begin{align}
    \sigma(pp \rightarrow hh)^{\rm up}_{13.6\, \rm TeV} &= 35.0\,\mathrm{fb}\times
         [1 - 0.885(\Delta\kappa_3) + 1.44\cdot 10^{-3}(\Delta\kappa_4) 
        + 0.342(\Delta\kappa_3)^2 \nonumber \\ & + 7.78\cdot 10^{-4}(\Delta\kappa_3\Delta\kappa_4) + 2.68\cdot 10^{-5}(\Delta\kappa_4)^2
         -1.58\cdot 10^{-3}(\Delta\kappa_3)^2(\Delta\kappa_4) \nonumber \\
         &  -1.87\cdot 10^{-5}(\Delta\kappa_3)(\Delta\kappa_4)^2 
         + 9.69\cdot 10^{-6}(\Delta\kappa_3)^2(\Delta\kappa_4)^2 ]\, ,
    \end{align}
    
\begin{align}
    \sigma(pp \rightarrow hh)^{\rm low}_{13.6\, \rm TeV}&= 26.9\,\mathrm{fb}\times
         [1 - 0.861(\Delta\kappa_3) + 1.49\cdot 10^{-3}(\Delta\kappa_4)
        + 0.323(\Delta\kappa_3)^2 \nonumber \\ & + 7.95\cdot 10^{-4}(\Delta\kappa_3\Delta\kappa_4) + 2.76\cdot 10^{-5}(\Delta\kappa_4)^2 
         -1.60\cdot 10^{-3}(\Delta\kappa_3)^2(\Delta\kappa_4) \nonumber \\ & -1.92\cdot 10^{-5}(\Delta\kappa_3)(\Delta\kappa_4)^2
         + 9.89\cdot 10^{-6}(\Delta\kappa_3)^2(\Delta\kappa_4)^2 ]\, ,
    \end{align}

\begin{align}
  \sigma(pp \rightarrow hh)^{\rm central}_{14\, \rm TeV} &= 32.9\,\mathrm{fb}\times
        [1 - 0.867(\Delta\kappa_3) + 1.48\cdot 10^{-3}(\Delta\kappa_4)
        + 0.329(\Delta\kappa_3)^2 \nonumber \\ &+ 7.80\cdot 10^{-4}(\Delta\kappa_3\Delta\kappa_4) + 2.73\cdot 10^{-5}(\Delta\kappa_4)^2 
         -1.57\cdot 10^{-3}(\Delta\kappa_3)^2(\Delta\kappa_4) \nonumber \\ & -1.90\cdot 10^{-5}(\Delta\kappa_3)(\Delta\kappa_4)^2 
        + 9.74\cdot 10^{-6}(\Delta\kappa_3)^2(\Delta\kappa_4)^2 ]\, ,
    \end{align}

\begin{align}
    \sigma(pp \rightarrow hh)^{\rm up}_{14\, \rm TeV} &= 37.3\,\mathrm{fb}\times
         [1 - 0.882(\Delta\kappa_3) + 1.45\cdot 10^{-3}(\Delta\kappa_4)
        + 0.341(\Delta\kappa_3)^2 \nonumber \\ & + 7.74\cdot 10^{-4}(\Delta\kappa_3\Delta\kappa_4) + 2.68\cdot 10^{-5}(\Delta\kappa_4)^2 
         -1.57\cdot 10^{-3}(\Delta\kappa_3)^2(\Delta\kappa_4) \nonumber \\ & -1.87\cdot 10^{-5}(\Delta\kappa_3)(\Delta\kappa_4)^2
         + 9.67\cdot 10^{-6}(\Delta\kappa_3)^2(\Delta\kappa_4)^2 ]\, ,
    \end{align}

\begin{align}
   \sigma(pp \rightarrow hh)^{\rm low}_{14\, \rm TeV} &= 28.8\,\mathrm{fb}\times
         [1 - 0.859(\Delta\kappa_3) + 1.50\cdot 10^{-3}(\Delta\kappa_4)
        + 0.321(\Delta\kappa_3)^2 \nonumber \\ & + 7.90\cdot 10^{-4}(\Delta\kappa_3\Delta\kappa_4) + 2.76\cdot 10^{-5}(\Delta\kappa_4)^2 
         -1.59\cdot 10^{-3}(\Delta\kappa_3)^2(\Delta\kappa_4) \nonumber \\ & -1.92\cdot 10^{-5}(\Delta\kappa_3)(\Delta\kappa_4)^2 
         + 9.84\cdot 10^{-6}(\Delta\kappa_3)^2(\Delta\kappa_4)^2 ]\, .
    \end{align}

 \begin{figure}[t]
\begin{center}\vspace{-0.2cm}
\begin{tabular}{cc}
\includegraphics[width=\textwidth]{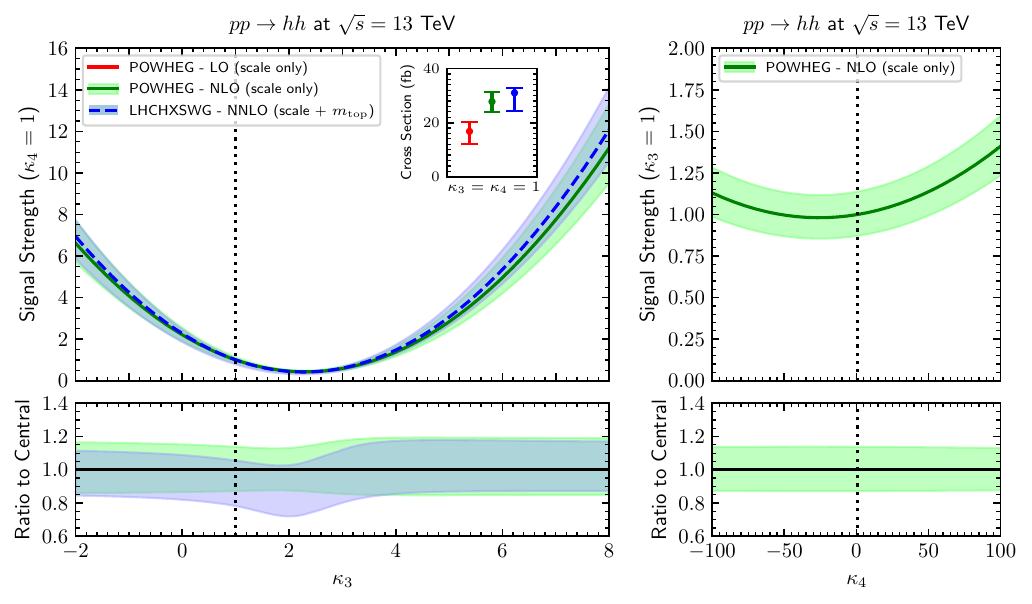}
&
\end{tabular}
\caption{\label{fig:13} Signal strength as a function of $\kappa_3$ (left panel) for $\kappa_4 = 1$ and as a function of $\kappa_4$ (right panel) for $\kappa_3 = 1$ at NLO accuracy in the full theory (green, solid). In the left panel we also show the FT$_{\rm approx}$ NNLO result for comparison (blue, dashed). The SM cross-sections are displayed in the inset, where also the LO result is included. The shaded uncertainty bands correspond to the 7-point QCD scale variation. For the FT$_{\rm approx}$ NNLO calculation, a component arising from the uncertainty in the choice of renormalisation scheme and of $m_{\mathrm{top}}$ is additionally included.}
\end{center}
\end{figure}

In Fig.~\ref{fig:13} we plot the signal strength for 13 TeV as a function of $\kappa_3$ for $\kappa_4 = 1$ (left panel) and as a function of $\kappa_4$ for $\kappa_3 = 1$ (right panel). The signal strength depends rather weakly on the value of $\kappa_4$, as one may expect since double Higgs production probes the quartic Higgs self-coupling only indirectly. The scale uncertainties of the NLO result in the full theory are at the 15\% level as evident from the two lower panels. In the left panel we also display the current recommendation of the Higgs cross section working group for the $HH$ cross-section as a function of $\kappa_3$. These predictions were obtained at NNLO accuracy in the so-called Full Theory (FT) approximation by rescaling them to the SM FT$_{\rm approx}$ NNLO result~\cite{Grazzini:2018bsd}, see Ref.~\cite{Amoroso:2020lgh} for additional details. The uncertainties are computed by probing three relative variations of $\mu_R = \mu_F \in \{ (1/2,1/2),(1,1),(2,2)\}$ and rescaled to match the conservative prescription of Ref.~\cite{Baglio:2020wgt} for the SM result, which combines the uncertainties arising from the choice of renormalisation scheme and scale of the top-quark mass with the $\mu_R$, $\mu_F$ scale uncertainties. We observe that the NLO predictions in the full theory nicely overlap with the FT$_{\rm approx}$ NNLO ones, both for the SM result and for values of $\kappa_3 \neq 1$. For reference, we also show the LO and the NLO cross section in the full theory alongside the NNLO FT$_{\rm approx}$ cross section in the inset of the left panel. Also in this case we observe that the NLO result overlaps with the NNLO FT$_{\rm approx}$ cross section.

\begin{figure}[t]
\begin{center}\vspace{-0.2cm}
\begin{tabular}{cc}
\includegraphics[width=0.49\textwidth]{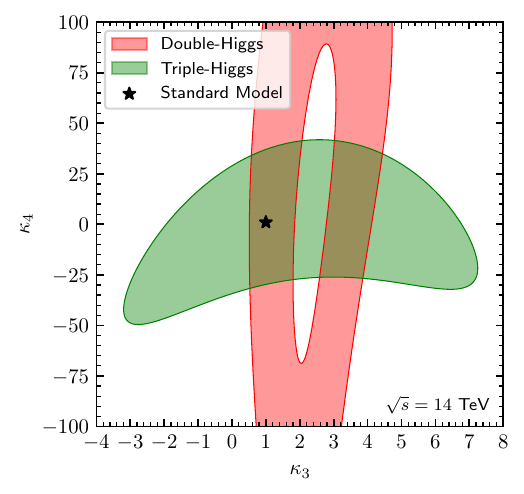}
&
\end{tabular}
\caption{\label{fig:14} Hypothetical constraints in the $\kappa_3-\kappa_4$ plane arising from inclusive double- (red) and triple- (green) Higgs production for HL-LHC at 14 TeV. The constraints are obtained assuming a 50\% uncertainty on the signal strength for double-Higgs production and an upper limit of 20 times the SM value for triple Higgs production.}
\end{center}
\end{figure}

Finally, in Fig.~\ref{fig:14} we display hypothetical constraints on $\kappa_3$ and $\kappa_4$ at the HL-LHC at 14 TeV arising from double-Higgs production assuming a 50\% uncertainty on the value of the signal strength $\mu_{HH}$. We also display hypothetical limits arising from triple-Higgs production, by assuming that the HL-LHC could set an $\mathcal O(20)$ bound on the triple-Higgs signal strength $\mu_{HHH}$. The predictions for triple-Higgs production have been obtained at LO using a private version of \texttt{MadGraph5\_aMC@NLO}~\cite{Alwall:2014hca}.  

In conclusion, in this Addendum  we have presented results for double-Higgs production at the LHC and at the HL-LHC including variations of the trilinear and of the quartic Higgs boson self-couplings.
We have performed a first exploratory study by considering constraints on $\kappa_4$ arising solely from inclusive double-Higgs production. 
A more refined analysis that includes also complementary constraints from of kinematic distribution would be necessary to assess the full potential of the HL-LHC to provide first bounds on the quartic Higgs self-coupling.
Our calculation is implemented in the POWHEG-BOX framework and is publicly available at the POWHEG-BOX webpage~\cite{PBOX}. We expect this implementation to be useful for the LHC experiments in conducting detailed sensitivity studies.

\acknowledgments

We are indebted to Javier Mazzitelli and Carlo Pandini for various discussions. We thank Franziska Rauscher for preliminary studies in the initial stages of this work. L.R. has been supported by the SNSF under contract PZ00P2 201878. P.W acknowledges support from a Grainger Fellowship at the University of Chicago.

\bibliographystyle{apsrevmod}
\bibliography{h4}

\end{document}